# MODELING PROTON- AND LIGHT ION-INDUCED REACTIONS AT LOW ENERGIES IN THE MARS15 CODE[*][§]

*I.L. Rakhno[#], N.V. Mokhov*

*Fermilab, Batavia, IL 60510, USA*

*K.K. Gudima*

*Institute of Applied Physics, National Academy of Sciences, Cisineu, Moldova*

## Abstract

An implementation of both ALICE code and TENDL evaluated nuclear data library in order to describe nuclear reactions induced by low-energy projectiles in the Monte Carlo code MARS15 is presented. Comparisons between results of modeling and experimental data on reaction cross sections and secondary particle distributions are shown.

[*]Work supported by Fermi Research Alliance, LLC, under contract No. DE-AC02-07CH11359 with the U.S. Department of Energy.
[§]Presented at the 6[th] International Particle Accelerator Conference (IPAC´15), May 3-8, 2015, Richmond, VA, USA.
[#]rakhno@fnal.gov

# MODELING PROTON- AND LIGHT ION-INDUCED REACTIONS AT LOW ENERGIES IN THE MARS15 CODE*

*I.L. Rakhno[#], N.V. Mokhov, Fermilab, Batavia, IL 60510, USA*
*K.K. Gudima, Institute of Applied Physics, National Academy of Sciences, Cisineu, Moldova*

## Abstract

An implementation of both ALICE code and TENDL evaluated nuclear data library in order to describe nuclear reactions induced by low-energy projectiles in the Monte Carlo code MARS15 is presented. Comparisons between results of modeling and experimental data on reaction cross sections and secondary particle distributions are shown.

## INTRODUCTION

Correct prediction of secondary particles, both neutral and charged ones, generated in proton-nucleus interactions below a few tens of MeV is required for various applications. The latter include, e.g., radiation studies for front-end of many proton accelerators, energy deposition studies for detectors, residual activation and radiation damage calculations, etc. In medical accelerators, low energy primary proton or deuteron beams are used either directly or as a tool to generate secondary neutron beams. Cascade models of various flavours fail to properly describe this energy region (see Fig. 1). Therefore, we opted to use the TENDL library developed by the Nuclear Research and Consultancy Group [1]. The evaluated data is provided in the ENDF/B format in the projectile energy range from 1 to 200 MeV, and the library is annually updated since 2008. In addition, a much more time-consuming approach utilized in a modified code ALICE [2] was also looked at. For both the options, the energy and angle distributions of all secondary particles are described with the Kalbach-Mann systematics. The following secondaries are taken into account: gammas, neutrons, protons, deuterons, tritons, $^3$He and $^4$He. The energy and angular distributions of all generated residual nuclei—including unstable ones—are accounted for as well.

## DETAILS OF THE IMPLEMENTATION AND FORMALISM

### TENDL Library

The TALYS-based evaluated nuclear data library (TENDL) contains data for direct use in both basic physics and applications. The evaluations were performed for practically entire periodic table except for hydrogen and helium. The library contains data for both stable and unstable target nuclei—all isotopes which live longer than 1 second were taken into account. At present, the list includes about 2400 isotopes. In fact, the TENDL library contains data not only for protons as projectiles, but also for light ions (d, t, $^3$He, $^4$He), neutrons and gammas. In current implementation, the data for protons and light ions are used when modelling low-energy inelastic reactions in MARS15 code [4-5]. For every interaction the following information is extracted from the library: (i) total inelastic cross section; (ii) energy and angular distributions of all above mentioned secondary particles and residual nuclei; (iii) yields of all the secondaries.

*Work supported by Fermi Research Alliance, LLC, under contract No. DE-AC02-07CH11359 with the U.S. Department of Energy
#rakhno@fnal.gov

The TENDL library is a collection of files in both ENDF/B and ACE format. We opted to use the source—ENDF/B format—because it is more easily readable which is a very important feature at the development and debugging stages. At the same time, there is little difference between the two formats with respect to computer memory requirements.

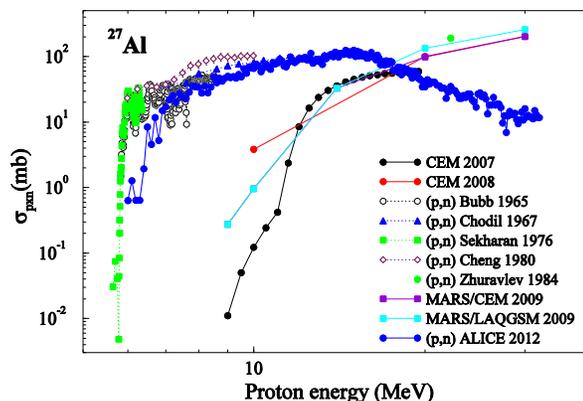

Figure 1: Calculated and measured [3] neutron production cross section on aluminum at low energies.

### ALICE Code

An alternative approach—using an event generator—was explored as well. For this purpose, the nuclear model code ALICE [2] based on a hybrid model of pre-compound decay, Weisskopf-Ewing evaporation and Bohr-Wheeler fission models was employed. It was re-designed in order to be used as an event generator for nucleon, photon and heavy-ion nuclear reactions at incident energies from 1 to 20-30 MeV matching CEM and LAQGSM at energies above 20-30 MeV in MARS15. At present, this option looks much more time consuming, but potentially it can offer some advantages for applications where accuracy of the full exclusive modeling is of major importance, e.g. when modeling a detector performance.

### Kalbach-Mann Systematics

Nowadays, continuum energy-angle distributions of secondary particles generated in low energy interactions are described usually with the Kalbach-Mann systematics [6] represented by the following equation:

$$f(\mu_b, E_a, E_b) = 0.5 * f_0(E_a, E_b)$$
$$\times \left[ \frac{a}{\sinh(a)} \left[ \cosh(a\mu_b) + r(E_a, E_b) \sinh(a\mu_b) \right] \right], \quad (1)$$

where $a = a(E_a, E_b)$ is a parametrized function, $r(E_a, E_b)$ is the pre-compound fraction as given by the evaluator, $\mu_b$ is cosine of scattering angle, $E_a$ and $E_b$ are energies of incident projectile and emitted particle, respectively, and $f_0(E_a, E_b)$ is total probability of scattering from $E_a$ to $E_b$ integrated over all angles. The function $a(E_a, E_b)$ depends mostly on the emission energy, and there is also a slight dependence on particle type and the incident energy at higher values of $E_a$.

### Coding Details

The corresponding processing and modeling software was written in C++ which provides substantial flexibility with respect to the computer memory used. In addition, the initialization of required evaluated data is performed dynamically whenever the modeling code encounters a nuclide not accounted for yet. The latter feature enables us to significantly reduce the amount of memory needed for extended systems with large number of materials.

## COMPARISONS WITH EXPERIMENTS

### Reaction Cross Section

Comparisons between calculated and measured reaction cross sections for several light and medium target nuclei important from a practical standpoint are shown for incident protons and deuterons in Figs. 2-4. The overall agreement between the predicted and measured cross sections is pretty good. The case of niobium is especially important because nowadays superconducting RF cavities are made of pure $^{93}$Nb. Fig. 4 also shows quality of prediction for generation of residual nuclei which are formed when emission of several light ejectiles is completed.

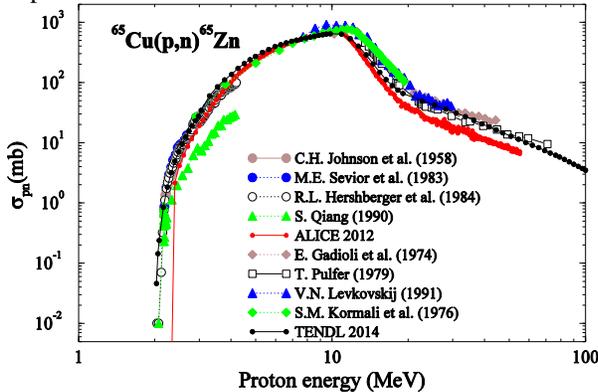

Figure 2: Calculated and measured [3] single neutron production cross section on $^{65}$Cu.

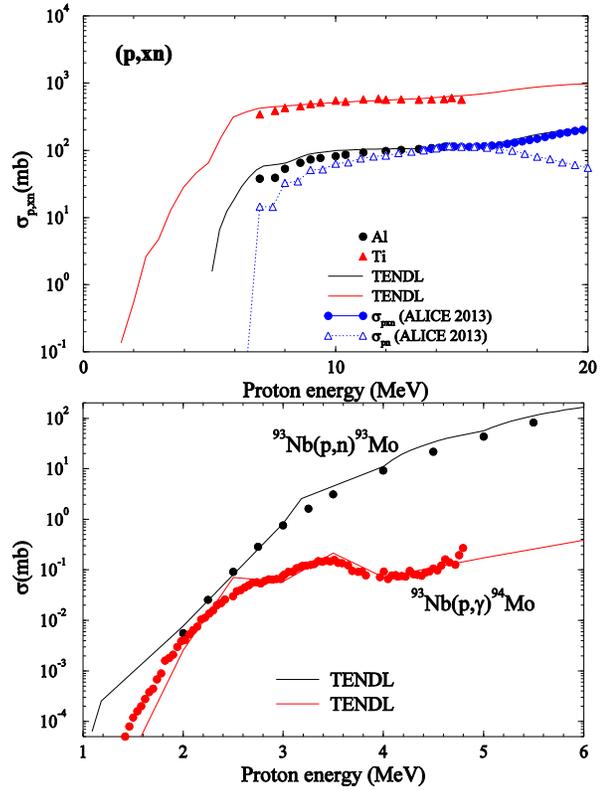

Figure 3: Calculated (lines) and measured (symbols) [3] neutron and gamma production cross sections for low energy protons. The data for (p,n) and (p,γ) reactions on $^{93}$Nb are from Refs. [7] and [8], respectively.

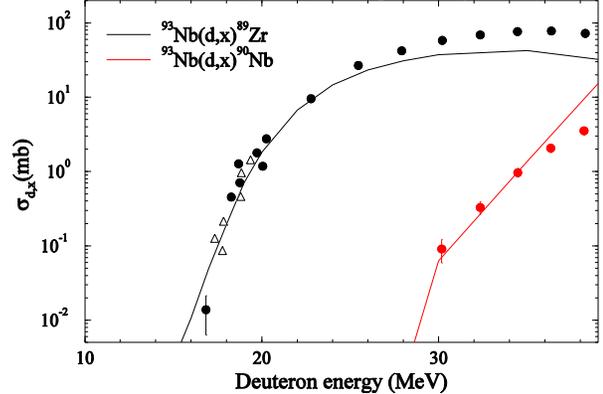

Figure 4: Calculated with TENDL (lines) and measured (symbols) deuteron-induced reaction cross sections. The data shown as circles and triangles are from Refs. [9] and [10], respectively.

### Energy Spectra of Secondary Particles

Energy spectra of secondary neutrons generated on light, medium and heavy target nuclei are shown in Fig. 5. At low secondary energies—where the major neutron emission takes place—the TENDL library provides reliable description of the energy distributions. At the very low projectile energies—10 MeV and lower—the library does not reproduce well some nuclear structure effects, but for high energy applications the agreement with the experimental data is quite adequate.

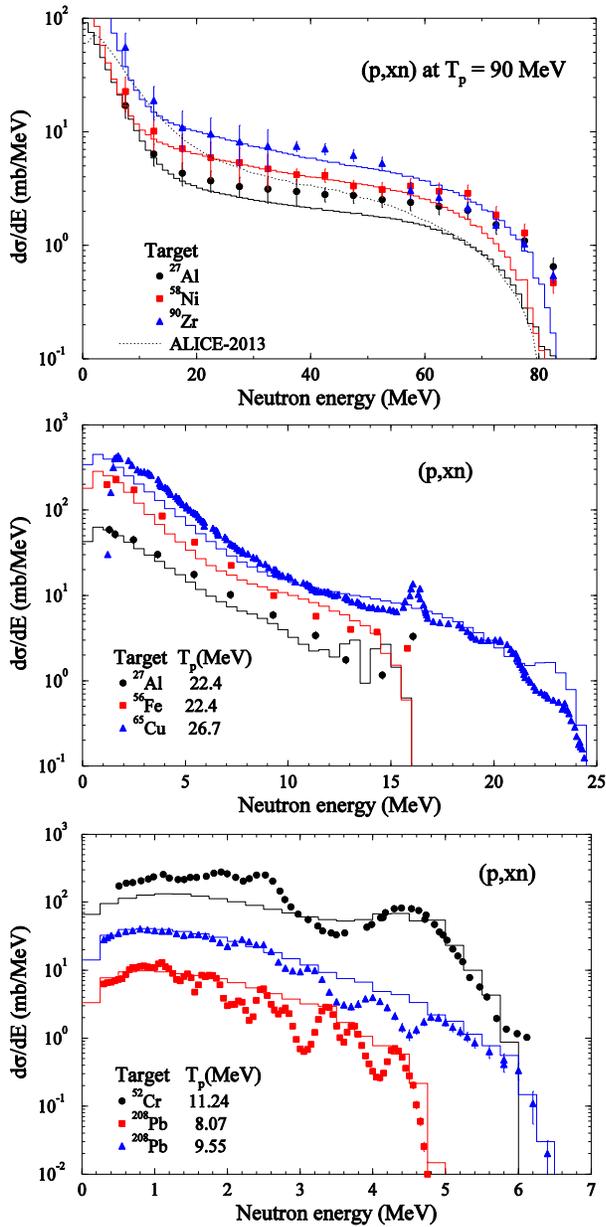

Figure 5: Calculated (solid histograms - TENDL) and measured (symbols) [3] energy spectra of secondary neutrons generated by protons on various target nuclei.

## Angular Distributions of Secondary Particles

Angular distributions of secondary neutrons and other light particles are shown in Fig. 6. One can see that both TENDL library and the ALICE-based generator can reproduce well the general features of the angular distributions: almost isotropic distributions at low emitted energies and forward-peaked distributions at higher secondary energies. The numerical agreement with the data for these target nuclei is also good. The transition region between low projectile energies described with TENDL library and higher energies can be broad because the library provides a good description of experimental data up to projectile energy of 200 MeV. At the same time, the transition region is expected to be dependent on target mass number because quality of the Kalbach-Mann systematics gets worse for light target nuclei.

The next step in these studies is thorough benchmarking of the TENDL implementation in the MARS15 code against macroscopic experimental data.

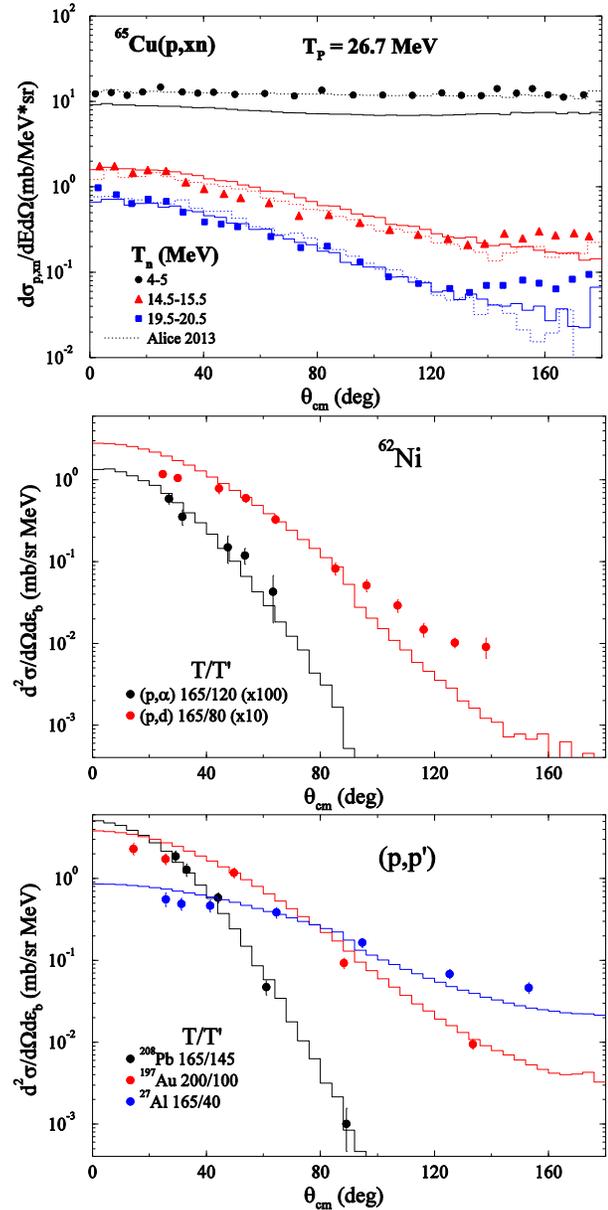

Figure 6: Calculated (solid histograms - TENDL) and measured (symbols) [3] angular distributions of secondary neutrons and other light particles generated by incident protons of various energies on various target nuclei. In the middle and lower parts of the Figure the projectile/ejectile energies, T/T´, are given in MeV.